\documentstyle[12pt]{article}
\begin{document}
\begin{center}
\Large {\bf  Non-Forward BFKL  at Next-to-Leading\\ Approximation
\footnote{Work supported in part by INTAS and in part by the
Russian Fund of Basic Researches.}}
\end{center}
\vskip 0.2cm \centerline{  V. Fadin \footnote{The speaker thanks
 R.~Fiore, M.~Kotsky and A.~Papa, his collaborators in works on
 which this talk is based.}}
\vskip 0.2cm \centerline{ Budker Institute for Nuclear Physics
and} \centerline{Novosibirsk State University, 630090
Novosibirsk, Russia}
\begin{abstract}
Representation of non-forward scattering amplitudes in the BFKL
approach is discussed and the results  obtained in the next-to
leading order  are briefly reviewed.
\end{abstract}
\vskip 0.2cm

\section{Introduction}
The BFKL approach \cite{BFKL} gives the most common basis for the
description of processes at large centre-of-mass energy $\sqrt s$
and fixed (not growing with $s$) momentum  transfers
$\sqrt{-t_i}$ in the framework of  perturbative QCD. This
approach was developed originally in the leading logarithmic
approximation (LLA), which means resummation of all terms of the
type $[\alpha _{s}\ln s]^{n}$ ($\alpha _{s}=g^2/(4\pi)$ is the
QCD coupling constant). Such approximation leads to a sharp
increase of cross sections with $s$:
\begin{equation}
{\sigma}^{LLA}_{tot}\sim {s^{\omega^B_{P}}\over
{\sqrt{\ln{s}}}}~, \,\,\,\,{\omega^B_{P}} = {g^2\over
\pi^2}N{\ln{2}} \label{a2}\label{a1}~,
\end{equation}
where $\omega^B_{P}$ is the LLA position of the most right
singularity in the complex momentum plane  of the  $t$-channel
partial wave with vacuum quantum numbers  (Pomeron singularity),
$N$ is the number of colours ($N=3$ for QCD). The Froissart bound
$\sigma_{tot}< C(\ln s)^2$ is violated in LLA, so that LLA can
not be applied at asymptotically large energies. The problem of
unitarization  of LLA results is extremely important from the
theoretical point of view. But in the region of energies
accessible for modern experiments it seems that the most
important disadvantage of LLA is that neither the scale of $s$
nor the argument of the running coupling constant $\alpha_s$ are
fixed in this approximation. To remove these uncertainties we
have to know radiative corrections to  LLA.

The calculation of the radiative corrections  has taken many
years of a hard work. Two years ago the kernel of the BFKL
equation was obtained in the next-to-leading order (NLO)
~\cite{FL98,CC} for the case of the forward scattering, i.e. for
momentum transfer $t=0$ and the singlet colour representations
the $t$-channel. In the $\overline{MS}$ renormalization scheme
with a reasonable scale setting the corrections appear to be
large, although they are not so large in non-Abelian physical
renormalization schemes with  BLM scale setting \cite{BFKLP}. The
problem of large corrections was widely discussed in literature
(see, for instance~\cite{papers}). Several ways to cure the
problem were suggested which provide an opportunity for
application of  the BFKL approach at NLO to high energy
phenomenology.

An important way of development of the BFKL approach became the
generalization of the obtained results to the non-forward
scattering~\cite{FF98}. For singlet colour representation in the
$t$-channel, the generalized approach can be used directly for
the description of a wide circle of physical processes, in
particular, diffraction phenomena.  The generalization for
non-singlet colour states is also of great importance, especially
for the antisymmetric colour octet state of two Reggeized gluons
in the $t$-channel. This case is especially important since the
BFKL approach is based on the gluon Reggeization, although the
Reggeization is not proved at NLO. Therefore, at this order the
Reggeization is a hypothesis which must be carefully checked.
This can be done using the "bootstrap" equations~\cite{FF98}
appearing from the requirement of the compatibility of the gluon
Reggeization with the $s$-channel unitarity.

In the next Section we remind  the representation of scattering
amplitudes in the BFKL approach. In Section 3 we discuss the
results obtained for the non-forward BFKL kernel and present some
of them. In Section 4 the same is done for the impact factors. The
problem of "bootstrap" is discussed in Section 5.

\section{Scattering amplitudes in the BFKL approach}

The BFKL approach is based on  the remarkable property of QCD -
the gluon Reggeization. Remind that we use the notion
"Reggeization"  in a strong sense \cite{FF98}. For the process
$A+B$ $\rightarrow A^{\prime }+B^{\prime }$ at large $s$ and
fixed $t\,\, ( s=(p_{A}+p_{B})^{2}\, , \, t=q^{2}\, , \,
 q=p_{A}-p_{A^{\prime }})$
it means that the  part of the amplitude corresponding to the
gluon quantum numbers (colour octet state and negative signature)
in the $t$-channel has the Regge form
\begin{equation}
({\cal A}_{8^-})_{AB}^{A^{\prime }B^{\prime }} = \Gamma
_{A^{\prime }A}^{c}(q) \left[ \left( \frac{-s}{-t}\right)
^{j(t)}-\left( \frac{s}{-t}\right) ^{j(t)} \right] \Gamma
_{B^{\prime }B}^{c}(q)\, ,  \label{z3}
\end{equation}
where $c$ is a color index  and $j(t)$ is the Regge trajectory
passing through zero at $t=1$. Therefore the gluon in QCD lies on
the Regge trajectory $j(t)=1+\omega(t)$ with $\omega(0)=0$, so
that it appears to be not an elementary object (as well as quark,
see \cite{FS}). $\Gamma _{P^{\prime }P}^{c}$ in (\ref{z3}) are
the particle-particle-Reggeon (PPR) vertices which do not depend
on $s$. Notice that  (\ref{z3}) represents correctly the
analytical structure of the scattering amplitude, which is quite
simple in this case.

The amplitudes with quantum numbers in the $t$- channel different
from the gluon ones are obtained in the BFKL approach using the
unitarity relations. For this purpose one needs to know inelastic
amplitudes. The main contributions in the unitarity relations
come from  the multi-Regge kinematics, where all produced
particles are strongly ordered in the rapidity space and all
their transverse momenta are fixed (not growing with $s$). In LLA
only such kinematics does contribute. In the next-to-leading
approximation (NLA), where terms with one extra power of
$\alpha_s$ compared to LLA are  kept,  production of a pair of
particles with rapidities of the same order is also admitted.

The production amplitudes have a complicated analytical structure
even in the multi-Regge kinematics. Fortunately,  we need to know
only real parts of these amplitudes. Due to the gluon
Reggeization they have a simple factorized form and can be
presented as
\begin{equation}
{\cal A}_{AB}^{\tilde{A}\tilde{B}+n} = 2s\Gamma
_{\tilde{A}A}^{c_{1}}(q_1) \left[\prod_{i=1}^{n}\frac{1}{t_{i}}
\gamma_{c_{i}c_{i+1}}^{P_{i}}(q_{i},q_{i+1})\left( \frac{s_{i}}
{\sqrt{\vec{k}_{i-1}^{2}\vec{k}_{i}^{2}}}\right)^{\omega
(t_{i})}\right]
\frac{1}{t_{n+1}}\left( \frac{s_{n+1}}{\sqrt{\vec{k}_{n}^{2}
\vec{k}_{n+1}^{2}}}\right) ^{\omega (t_{n+1})} \Gamma
_{\tilde{B}B}^{c_{n+1}}(q_{n+1})\ , \label{z11}
\end{equation}
where $\gamma _{c_{i}c_{i+1}}^{P_{i}}(q_{i},q_{i+1})$ are the
so-called Reggeon-Reggeon-particle (RRP) vertices, i.e. the
effective vertices for the production of the particles $P_i$ with
momenta $k_{i}$=$q_{i}-q_{i+1}$ in the collision of  Reggeons
with momenta $q_{i}$ and $-q_{i+1}$ and colour indices $c_{i}$ and
$c_{i+1}$, $q_0\equiv p_A, q_{n+1}\equiv -p_B,\,\,
s_i=(k_{i-1}+k_i)^2,\, k_0\equiv p_{\tilde{A}}, k_{n+1}\equiv
p_{\tilde{B}}$, vector sign is used for denotation of components
of momenta transverse to the plane of the initial momenta $p_A,
p_B$. Pay attention that we have taken definite energy scales in
the Regge factors in (\ref{z11}) as well as in (\ref{z3}). In
principle, we could take an arbitrary scale $s_{R}$ \cite{rio98};
in this case the PPR and RRP vertices would become dependent on
$s_{R}$. Of course, physical results do not depend on the scale.

In  LLA only one particle can be produced in each  RRP vertex, and
since our Reggeons are Reggeized gluons, this particle can be
only a gluon. The situation changes in  NLA, where one of the
vertices  can represent production of a pair of particles with
fixed (not increasing with $s$) invariant mass. We can treat this
kinematics using the effective vertices $\gamma
_{cc^{\prime}}^{GG}$ and $\gamma _{cc^{\prime}}^{Q\bar Q}$  for
production of two gluons and a quark-antiquark pair  in
Reggeon-Reggeon collisions. Analogously, we use  effective
vertices $\Gamma_{\{(f)\}P}^{c}$ for production of states
$\{(f)\}$ containing an extra particle in the fragmentation
region of the particle $P$ in the process of scattering of this
particle off the Reggeon.

In the BFKL approach Pomeron appears as a compound state of two
Reggeized gluons in the $t$- channel in the colour singlet
representation. In general case, the amplitude for the high energy
process $A+B\rightarrow A^{\prime }+B^{\prime }$ at fixed momentum
transfer $\sqrt{-t}$ is presented in this approach as the
convolution of  the impact factors  $\Phi _{A^{\prime}A}$ and
$\Phi _{B^{\prime}B}$,  which describe the $A \rightarrow
A^\prime$ and $B \rightarrow B^\prime$  transitions in the
particle-Reggeon scattering processes, and the Green's function
$G$  for Reggeon-Reggeon scattering \cite{FF98}:
\[
{\cal A} _{AB}^{A^{\prime }B^{\prime }}
=
\frac{is}{{\left( 2\pi \right) ^{D-1}}}\int \frac{d^{D-2}q_{1}}{\vec{q}%
_{1}^{~2} \vec{q}_{1}^{\, \prime\, 2}}\int \frac{d^{D-2}q_{2}}{\vec{%
q}_{2}^{~
2} \vec{q}_{2}^{\, \prime \, 2}}\,\int_{-i\infty +\delta}^{i\infty +\delta} \frac{d\omega}{\mbox{sin}(\pi\omega)}\sum_{ {\cal R},\nu} \Phi _{A^{\prime
}A}^{\left( {\cal R},\nu \right) }\left( \vec{q}_{1};\vec{q}%
;s_{0}\right)
\]
\begin{equation}
\times
\left[ \left( \frac{-s}
{s_{0}}\right)
^{\omega}-\tau\left( \frac{s}{s_{0}}\right)^{\omega}\right]{ G_{\omega }^{\left( {\cal R}\right) }\left( \vec{q}_{1},\vec{q}%
_{2},\vec{q}\right)}  \Phi _{B^{\prime }B}^{\left( {\cal R},%
\nu \right) }\left( -\vec{q}_{2};-\vec{q};s_{0}\right)\,.
\label{5}
\end{equation}
Here and below  $q_i^\prime \equiv q_i - q$, $\vec {q}_i$ are
$D-2$ dimensional vectors ($D=4+2\epsilon$ is the space-time
dimension different from $4$ to regularize the infrared
singularities), $q \simeq \vec q \,$ is the momentum transfer in
the process $A + B \rightarrow A^\prime + B^\prime$.  The sum in
(\ref{5}) is taken over irreducible representations  ${\cal R}$
of the colour group, which are contained in the product of two
adjoint representations,  and over states  $\nu$  from  full sets
of states for these representations; $\tau$ is the signature equal
$+1 (-1)$ for symmetric (antisymmetric) representations and
$s_{0}$ is an energy scale.     $\Phi _{A^{\prime }A}^{\left(
{\cal R},\nu \right) }$  and  $\Phi _{B^{\prime }B}^{\left( {\cal
R},\nu \right) }$ are the projected on the states $\left( {\cal
R},\nu \right)$ impact factors, and $G_\omega^{\left( {\cal R}
\right)}$ are the Mellin transforms of the Green's functions for
Reggeon-Reggeon scattering. All dependence from $s$ is determined
by $G_\omega^{\left( {\cal R} \right)}$, which obey the
equation~\cite{FF98}:
\begin{equation}
\omega G_\omega ^{\left( {\cal R}\right) }\left( \vec q_1,\vec q_2,\vec
q\right) =
\vec q_1^{\:2} \vec q_1^{\:\prime\:2}\delta ^{\left( D-2\right)
}\left( \vec q_1-\vec q_2\right) +\int \frac{d^{D-2}r}{{\vec r}
_{}^{\,2}\vec r^{\:\prime\:2}}{\cal K}^{\left( {\cal R}
\right) }\left( \vec q_1,{\vec r};\vec q\right) G_\omega ^{\left( {\cal R}
\right) }\left( {\vec r},\vec q_2;\vec q\,\right) ~,    \label{6}
\end{equation}
where the  kernel is given by the sum of the "virtual" part,
defined by the gluon trajectory $j(t)=1+\omega \left( t\right) $,
and the "real" part ${\cal K}_r^{\left( {\cal R} \right) }$,
related to the real particle production in  Reggeon-Reggeon
collisions:
\begin{equation}
{\cal K}^{\left( {\cal R}\right)}\left( \vec q_1,\vec q_2;
\vec q\,\right) =\left[ \omega \left( -\vec q_1^{\:2}\right)
+\omega \left( - \vec q_1^{\:\prime\:2} \right)\right]
\vec q_1^{\:2}\vec q_1^{\:\prime\:2}\delta ^{\left( D-2\right) }
\left( \vec q_1-\vec q_2\right) +
{\cal K}_r^{\left( {\cal R}\right) }\left( \vec q_1,\vec q_2;\vec
q\right) ~.  \label{7}
\end{equation}
In LLA both the gluon trajectory and the "real"  part of the
kernel are pure gluonic:
\begin{equation}
\omega ^{\left( 1\right) }\left( t\right) =\frac{g^2Nt}{2\left(
2\pi \right) ^{D-1}}\int \frac{d^{D-2}q_1}{\vec
q_1^{\:2}\vec{q}_{1}^{\:\prime \:2}}
=-\frac{g^2N\Gamma(1-\epsilon)\Gamma^2(\epsilon)}
{(4\pi)^{D/2}\Gamma(2\epsilon)}\left( \vec
q^{\:2}\right)^{\epsilon} \label{8},
\end{equation}
where $\Gamma(x)$ is the Euler gamma-function, and
\begin{equation}
{\cal K}_r^{\left( {\cal R}\right) B}\left( \vec q_1,\vec q_2;\vec
q\,\right) =\frac{g^2c_R}{(2\pi )^{D-1}}\left( \frac{\vec q_1^{\:2}\vec
q_2^{\:\prime\: 2}+\vec q_2^{\:2}\vec q_1^{\:\prime\:2}}{(\vec q_1- \vec
q_2)^2}-\vec q^{\:2}\right) ~,  \label{9}
\end{equation}
where the superscript $B$ means the leading (Born) approximation
and the colour group coefficients $c_R$ for the most important
cases of the singlet $(R=1)$ and the antisymmetric octet $(R=8)$
representations are $c_1=N,\,\,\,c_8=\frac N2 ~.$

The  representation (\ref{5}) for the scattering amplitudes, as
well as the equation (\ref{6}) for the Green's function and the
form (\ref{7}) of the kernel, are valid in NLA as well as in LLA.
Both the kernel of the BFKL  equation and the impact factors are
unambiguously defined in terms of the gluon Regge trajectory and
the effective vertices for the Reggeon-particle
interactions~\cite{FF98}.

At the two-loop level the gluon trajectory was calculated in
~\cite{F}. It has both gluon and quark contributions.  In the
case of $n_f$ massless quark flavors it can be presented as
\begin{equation}
\label{18}\omega^{(2)}(t) = \frac{g^2t}{(2\pi)^{D-1}}
\int\frac{d^{(D-2)}q_1}{\vec q_1^{\:2}\vec q_1^{\:\prime\:2}}
\left[F(\vec q_1,\vec q)-2F(\vec q_1,\vec q_1)\right] ~,
\end{equation}
where
$$F(\vec q_1,\vec q)=\frac{2g^2Nn_f\Gamma
\left( 2-\frac D2\right) \Gamma ^2\left( \frac D2\right) }{(4\pi
)^{\frac D2}\Gamma \left( D\right)\left( \vec q^{\:2}\right)
^{(2-\frac D2)} } -\frac{g^2N^2\vec q^{\:2}}{4(2\pi)^{D-1}}
\int\frac{d^{(D-2)}q_2}{\vec q_2^{\:2}(\vec q_2-\vec q)^2}
$$
$$
\times \left[\ln{\left(\frac{\vec q^{\:2}}{(\vec q_1-\vec
q_2)^2}\right)}-2\psi(D
-3)-\psi\left(3-\frac{D}{2}\right)+2\psi\left(\frac{D}{2}-2\right)+
\psi(1)\right.
$$
\begin{equation}
\label{19} \left.
+\frac{2}{(D-3)(D-4)}+\frac{D-2}{4(D-1)(D-3)}\right]~,
\end{equation}
$\psi(x)=\Gamma^{\prime}(x)/\Gamma(x)~.$ Let us stress that we
use  perturbative expansion in terms of the bare coupling $g$
related to the renormalized coupling $g_\mu $ in the
${\overline{MS}}$ scheme by the relation
\begin{equation}
g=g_\mu \mu ^{-\mbox{\normalsize $\epsilon$}}\left[ 1+\left(
\frac{11}3-\frac 23\frac{n_f}N\right) \frac{\bar g_\mu
^2}{2\epsilon }\right] ~,\,\,\,\bar g_\mu ^2=\frac{g_\mu
^2N\Gamma (1-\epsilon )}{(4\pi )^{2+{\epsilon }}}~.  \label{20}
\end{equation}
In the limit ${\epsilon \rightarrow 0}$ we have~\cite{F}:
\[
\omega(t) = -\bar g^2_{\mu}\left(\frac{\vec q^{~2}}{\mu
^2}\right)^ {\epsilon} \frac{2}{\epsilon}\left\{1+\frac{\bar
g^2_{\mu}}{\epsilon}
\left[\left(\frac{11}{3}-\frac{2}{3}\frac{n_f}{N}%
\right) \left(1-\frac{\pi^2}{6}\epsilon ^2\right)-\right.\right.
\]
\[
\left.\left. \left(\frac{\vec q^{~2}}{\mu ^2}\right)^
{\epsilon}\left(\frac{11}{6}+\left(\frac{\pi^2}{6}-
\frac{67}{18}\right)
\epsilon+\left(\frac{202}{27}-\frac{11\pi^2}{18}-\zeta(3)\right)
\epsilon^2-\right.\right.\right.
\]
\begin{equation}
\left.\left.\left.
\frac{n_f}{3N}\left(1-\frac{5}{3}\epsilon+\left(
\frac{28%
}{9}-\frac{\pi^2}{3}\right)\epsilon^2\right)\right)\right]\right\}.
\label{r7}
\end{equation}

\section{Non-forward  BFKL kernel}

The general form of the kernel for arbitrary representation
${\cal{R}}$ of the colour group in the $t$-channel is given by
 (\ref{7}). The "virtual" part of the kernel is universal (does
not depend on ${\cal{R}}$) and is determined by the gluon Regge
trajectory.  The "real" part can be presented as~\cite{FF98}:
\[
{\cal K}_{r}^{\left( {\cal R}\right) }\left(
\vec{q}_{1},\vec{q}_{2}; \vec{q}\,\right) =\frac{\langle
c_1c_1^{\prime }|\hat {{\cal P}}_{ {\cal R}}|c_2c_2^{\prime
}\rangle }{2n_{\cal R}} \int_0^{\infty} \frac{ds_{_{RR}}}{\left(
2\pi \right)^{D}}\theta \left( s_{_{\Lambda }}-s_{_{RR}} \right)
\sum_{\left\{ f\right\}}\int \gamma _{c_1c_2}^{\left\{ f\right\}
}\left( q_1,q_2\right) \left( \gamma _{c_1^{\prime }c_2^{\prime
}}^{\left\{ f\right\} }\left( q_1^{\prime },q_2^{\prime }\right)
\right) ^{*}d\rho _f
\]
\begin{equation}
-\frac{1}{2}\int
\frac{d^{D-2}r}{{\vec{r}}^{\,2}\vec{r}^{\:\prime\:2}}{\cal
K}_{r}^{\left( {\cal R}\right) B}\left( \vec{q}
_{1},{\vec{r}};\vec{q}\right) {\cal K}_{r}^{({\cal
R}){\scriptsize {B}} }\left(
{\vec{r}},\,\vec{q}_{2};\,\vec{q}\,\right) \,\ln\left( \frac{
s_{_{\Lambda
}}^{2}}{({\vec{r}}-\vec{q}_{1})^{2}({\vec{r}}-\vec{q}_{2})^{2}}
\right) ~,   \label{11}
\end{equation}
where $\hat {{\cal P}}_{ {\cal R}}$ is the operator  for
projection of two-gluon colour state in the $t$-channel on the
representation $R$,  $n_{{\cal R}}$ is the number of independent
states in  ${\cal R}$, $q_{1}=\beta p_{A}+q_{1\perp }\ $ and $\
-q_{2}=\alpha p_{B}-q_{2\perp }$ are the Reggeon momenta, $
s_{_{RR}}=\left( q_{1}-q_{2}\right) ^{2}=s\alpha\beta - (\vec q_1
-\vec q_2)^2$, the sum $\left\{ f\right\} $ is performed over all
states $f$ which are produced in the Reggeon-Reggeon collisions
and over all discreet quantum numbers of these  states, $d\rho
_f$ is the phase space element for such  state,
\begin{equation}
d\rho _f=\frac{1}{n!}\left( 2\pi \right) ^D\delta ^{\left(
D\right) } (q_1-q_2- \mathop{\textstyle\sum}_{i\in f }k_i)
\prod_{i\in f } \frac{d^{D-1}k_i}{\left( 2\pi \right)
^{D-1}2\epsilon _i}\ ~,\label{13}
\end{equation}
$n$ is a number of identical particles in the  state $ f $ . The
second term in the r.h.s. of (\ref{11}) serves for the
subtraction of the contribution of the large $s_{_{RR}}$ region
in the first term, in order to avoid a double counting of this
region in the BFKL equation. The intermediate parameter
$s_{_{\Lambda }}$ in (\ref{11}) must be taken tending to
infinity. In  LLA only the one-gluon production does contribute
and (\ref{11}) gives for the kernel its LLA value (\ref{9}); in
NLA the contributing states include also two-gluon and
quark-antiquark states. At large $s_{_{RR}}$ only the
contribution of the two-gluon production does survive in the
first integral, so that the dependence on $s_{_{\Lambda }}$
disappears in (\ref{11}) due to the factorization property of the
two-gluon production vertex~ \cite{FF98}.

The remarkable properties of the kernel are
\begin{equation}
{\cal K}_{r}^{({\cal R}) }( 0,\vec{q}_{2};\vec{q}\,) = {\cal
K}_{r}^{({\cal R}) }( \vec{q}_{1},0;\vec{q}\,) ={\cal
K}_{r}^{({\cal R}) }( \vec{q},\vec{q}_{2};\vec{q} \,) ={\cal
K}_{r}^{({\cal  R}) }( \vec{q}_{1},\vec{q};\vec{q}\,) =0 ~,
 \label{16}
\end{equation}
and
\begin{equation}
{\cal K}_{r}^{({\cal R}) }( \vec{q}_{1},\vec{q}_{2};\vec{q} \,)=
{\cal K}_{r}^{({\cal  R}) }( \vec{q} _{1}^{\:
\prime},\vec{q}_{2}^{\: \prime };-\vec{q}\,) ={\cal
K}_{r}^{({\cal  R}) }( -\vec{q}_{2},-\vec{q}_{1};- \vec{q}\,) ~.
 \label{17}
\end{equation}
The properties (\ref{16}) mean that the kernel turns into zero at
zero transverse momenta of the Reggeons and  follow from the gauge
invariance; (\ref{17}) are  the consequences of the symmetry of
the r.h.s. of~(\ref{11}).

The most interesting representations ${\cal R}$ are the colour
singlet (Pomeron channel) and the antisymmetric colour octet
(gluon channel). We have for the singlet case
\begin{equation}
\langle c_1c_1^{\prime }|\hat {{\cal P}}_1|c_2c_2^{\prime
}\rangle =\frac{ \delta _{c_1c_1^{\prime }}\delta
_{c_2c_2^{\prime }}}{N^2-1}\ ,\quad n_1=1\ ~, \label{14}
\end{equation}
and for the antisymmetric octet case
\begin{equation}
\langle c_1c_1^{\prime }|\hat {{\cal P}}_{8-}|c_2c_2^{\prime
}\rangle =\frac{ f_{c_1c_1^{\prime }c}f_{c_2c_2^{\prime }c}}N\ ,\
\ \ \ n_8=N^2-1 ~, \label{15}
\end{equation}
where $f_{abc}$ are the structure constants of the colour group.

The quark contribution to the "real" part for any representation
$\cal R $  is given by a superposition of the Abelian and
non-Abelian pieces \cite{FFP}. The Abelian piece was calculated
many yes ago in the framework of Quantum Electrodynamics
\cite{cheng}, but, unfortunately, it has rather complicated form
and can not be presented here.  In the gluon channel this piece
is absent and the quark contribution in this channel has very
simple form:
\[
{\Large {\cal K}}{r}^{\left( 8-\right) Q}\left(
\vec{q}_{1},\vec{q}_{2}; \vec{q}\,\right)
=\frac{g^{4}n_{f}N}{(2\pi )^{D-1}} \frac{\Gamma (1-\epsilon
)}{\epsilon (4\pi )^{2+\epsilon }}\frac{\Gamma^2 (2+\epsilon
)]}{\Gamma (4+2\epsilon )}
 \left\{ \frac{({\vec{k}}_{{}}^{2})^{\epsilon }}{{\vec{k}}
^{2}}(\vec{q}_{1}^{\:2}\vec{q}_{2}^{\:\prime \:2}+\vec{q}_{2}^{\:2}\vec{q}
_{1}^{\:\prime \:2})+\vec q^{\:2} \left((\vec q^{\:2})^\epsilon\right. \right. \]
\begin{equation}
\left.\left. - (\vec{q} _{1}^{\:2})^\epsilon -
(\vec{q}_{2}^{\:2})^\epsilon \right)
 -\frac{(\vec{q}_{1}^{\:2}\vec{q}_{2}^{\:\prime \:2}-\vec{q}_{2}^{\:2}
\vec{q}_{1}^{\:\prime \:2})}{{\vec{k}}_{{}}^{2}} \left( (\vec{q}
_{1}^{\:2})^{\epsilon }-(\vec{q}_{2}^{\:2})^{\epsilon }\right) +(\vec{q}
_{1}\leftrightarrow \vec{q}_{1}^{\:\prime },\vec{q}
_{2}\leftrightarrow \vec{q}_{2}^{\:\prime })\right\},  \label{23}
\end{equation}
where $\vec k =\vec q_1-\vec q_2=\vec q_1^{\prime}-\vec
q_2^{\prime}$. The properties (\ref{16}),(\ref{17})  are evident
from (\ref{23}).

The gluon piece of the "real"  part of the kernel includes
contributions of the one-gluon and two-gluon productions. The
one-gluon contribution is determined by the RRG-vertex and
depends on the representation  $\cal R $ only through the
numerical coefficient $c_R$ in the same way  as in the Born case
(\ref{9}). The calculation of the RRG-vertex has a long history.
The quark part of the vertex was calculated at once  at arbitrary
$D$~\cite{FFQ94}. Instead, in the gluon part firstly only the
terms finite at $\epsilon\rightarrow0 $ were kept~\cite{FL93}.
But then it was understood that in order to find the  BFKL kernel
the RRG-vertex must be calculated at arbitrary $D$ in the region
of  small transverse momenta $\vec k$ of the produced gluon. It
was done in ~\cite{FFK96b}. Later the results
of~\cite{FL93,FFK96b} were obtained by another method
in~\cite{DDS}. But after all it appeared \cite{FFK00} that  for
the verification of the "bootstrap" equation the vertex  must be
known at arbitrary $D$ in a wider kinematical region. Therefore
it became clear that the vertex should be calculated  at
arbitrary $D$, especially taking into account that it can be used
not only for the check of the "bootstrap", but, for example, in
the Odderon problem at NLO and so on. Quite recently \cite{FFP00}
the vertex and corresponding contribution to the non-forward BFKL
kernel were calculated at arbitrary $D$.

The two-gluon contribution, analogously to the quark one, for any
representation  $\cal R $  is given  by a linear combination of
two pieces. Only one of them enters in the antisymmetric octet
kernel. It was calculated recently \cite{FG}. In the case of
arbitrary $D$ the integral representation was obtained.

We have not enough space here to show   neither the one-gluon nor
the two-gluon contributions at arbitrary $D$, so that we present
only the sum of these contributions to the "real" part of the
BFKL kernel for the antisymmetric octet representation in the
limit $\epsilon \rightarrow 0$ \cite{FG}:
\[
{\cal K}_r^{G\,(8-)}(\vec q_1,\vec q_2; \vec{q})
        =  \frac{g^2N}{2(2\pi)^{D-1}}
\left\{ \left( \frac{\vec{q}_1^{\:2} \vec{q}_2^{\:\prime\:2}
+ \vec{q}_1^{\:\prime\:2} \vec{q}_2^{\:2}}{\vec k ^{\:2}}-
\vec{q}^{\:2}\right)\right.
\]
\[
\times \left(1+ \frac{g^2N \Gamma(1-\epsilon)(\vec k ^{\:2})^\epsilon}
{(4\pi)^{2+\epsilon}}  \left(-\frac{11}{6\epsilon}+\frac{67}{18}
-\zeta(2)+\epsilon\left(
-\frac{202}{27} +7\zeta(3) +\frac{11}{6}\zeta(2)\right)\right)\right)
\]
\[
+\frac{g^2N \Gamma(1-\epsilon)}{(4\pi)^{2+\epsilon}}\left[\vec q^{\,2}
\left(\frac{11}{6}\ln\left(\frac{\vec q_1^{\:2}\vec q_2^{\:2}}
{\vec q^{\:2}\vec k^{\:2}}\right)
+\frac{1}{4}\ln\left(\frac{\vec q_1^{\:2}}{\vec q^{\:2}}\right)
\ln\left(\frac{\vec q_1^{\:\prime 2}}{\vec q^{\:2}}\right)
+\frac{1}{4}\ln\left(\frac{\vec q_2^{\:2}}{\vec q^{\:2}}\right)
\ln\left(\frac{\vec q_2^{\:\prime 2}}{\vec q^{\:2}}\right)\right)
\right.
\]
\[
+\frac{1}{4}\ln^2\left(\frac{\vec q_1^{\:2}}{\vec q_2^{\:2}}\right)
-\frac{\vec q_1^{ \:2}\vec q_2^{\:\prime \:2}+\vec q_2^{ \:2}
\vec q_1^{\:\prime \:2}}{2\vec k^{\:2}} \ln ^2\left(\frac
{\vec q_1^{\:2}}{\vec q_2^{ \:2}}
\right)
+\frac{\vec q_1^{ \:2}\vec q_2^{\:\prime \:2}-\vec q_2^{ \:2}
\vec q_1^{\:\prime \:2}}{\vec k^{\:2}} \ln \left(\frac{\vec q_1^{\:2}}
{\vec q_2^{ \:2}}
\right)\left(\frac{11}{6}-\frac{1}{4}\ln \left(\frac{\vec q_1^{ \:2}
\vec q_2^{ \:2}}{\vec k^{\:4}}\right)\right)
\]
\[
+\frac{1}{2}
[\vec q^{\:2}(\vec k^{\:2}-\vec q_1^{\:2}-\vec q_2^{ \:2})+2
\vec q_1^{ \:2}\vec q_2^{ \:2}-\vec q_1^{\:2}\vec q_2^{\:\prime \:2}-
\vec q_2^{ \:2}\vec q_1^{\:\prime \:2}+\frac{\vec q_1^{ \:2}
\vec q_2^{\:\prime \:2}-\vec q_2^{ \:2}\vec q_1^{\:\prime \:2}}
{\vec k^{\:2}}(\vec q_1^{\:2}-\vec q_2^{ \:2})]
\]
\begin{equation}
\left.\left.\times \int_0^1\frac{dx}{(\vec q_1(1-x)+\vec q_2 x)^2}\ln
\left( \frac {\vec q_1^{ \:2}(1-x)+\vec q_2^{ \:2}x}{\vec k^{\:2}x(1-x)}
\right)\right]\right\}
+\frac{g^2N}{2(2\pi)^{D-1}}\Biggl\{\vec q_i \longleftrightarrow
\vec q_i^{\:\prime}\Biggr\}~,
\label{24}
\end{equation}
where $\zeta(n)$ is the Riemann zeta-function. Note, that the the
one-gluon and the two-gluon contributions separately have
singularities $\sim 1/\epsilon^2$. Due to their cancellation  the
leading singularity of the kernel is $1/\epsilon$. It turns again
into $\sim 1/\epsilon^2$ after subsequent integrations of the
kernel because of the singular behaviour of the kernel at $\vec
k^{\:2}=0$. The additional singularity arises from the region of
small $\vec k^{\:2}$, where   $\epsilon |\ln \vec k^{\:2}|\sim
1$. Therefore we have not expanded  in $\epsilon$ the  term
$\left(\vec k^{\:2}\right)^{\epsilon}$. The terms $\sim
~\epsilon$  are taken into account in the coefficient of the
divergent at  $\vec k^{\:2}=0$ expression in order to conserve
all  giving nonvanishing in the limit $\epsilon\rightarrow 0$
contributions after the integrations.

The integral in (\ref{24}) can be presented in another form:
\[
\int_0^1\frac{dx}{(\vec q_1(1-x)+\vec q_2 x)^2}\ln \left( \frac{\vec q_1^{ \:2}(1-x)+\vec q_2^{ \:2}x}{\vec k^{\:2}x(1-x)}\right)=\int_0^\infty \frac{dz}{z+{\vec k^{\:2}}}\frac{1}{\sqrt{(\vec q_1^{ \:2}+\vec q_2^{ \:2}+z)^2-4\vec q_1^{ \:2}\vec q_2^{ \:2}}}
\]
\begin{equation}
\times \ln\left(\frac{\vec q_1^{ \:2}+\vec q_2^{ \:2}+z+\sqrt{(\vec q_1^{ \:2}+\vec q_2^{ \:2}+z)^2-4\vec q_1^{ \:2}\vec q_2^{ \:2}}}{{\vec q_1^{ \:2}+\vec q_2^{ \:2}+z-\sqrt{(\vec q_1^{ \:2}+\vec q_2^{ \:2}+z)^2-4\vec q_1^{ \:2}\vec q_2^{ \:2}}}}\right)~.
\label{25}
\end{equation}
It is possible also to express the integral in (\ref{24}) in terms of dilogarithms, but this expression is not very convenient:
\[
\int_0^1\frac{dx}{(\vec q_1(1-x)+\vec q_2 x)^2}\ln \left( \frac{\vec q_1^{ \:2}(1-x)+\vec q_2^{ \:2}x}{\vec k^{\:2}x(1-x)}\right)=-\frac{2}{|\vec q_1||\vec q_2|\sin \phi}
\]
\begin{equation}
\times \left[\ln \rho \arctan{\frac{\rho \sin \phi}{(1-\rho \cos \phi)}} +Im \left(L\left(\rho\exp {i\phi}\right)\right)    \right],
\label{26}
\end{equation}
where $\phi$ is the angle between $\vec q_1$ and $\vec q_2$,
\begin{equation}
\rho=min\left(\frac{|\vec q_1|}{|\vec q_2|},\frac{|\vec q_2|}{|\vec q_1|} \right), \,\,\,
L(z)=\int_0^z\frac{dt}{t}\ln(1-t).
\label{27}
\end{equation}
The  symmetries  (\ref{17}) of the kernel are easily seen. The first of them
is explicit  in (\ref{24}). To notice the second it is sufficient to
change $x\leftrightarrow (1-x)$ in the integral in (\ref{24}).

In order to check that the kernel (\ref{24}) turns into zero at
zero transverse momenta of  Reggeons   one has to know the
behaviour of the integral in (\ref{24}). A suitable for this
purpose representation is given in (\ref{25}). From this
representation one can  see that singularities of the integral at
zero transverse momenta of the Reggeons are not more than
logarithmic. After this no problems remain to verify (\ref{16}).

\section{Non-forward impact factors}
The impact factors can be expressed through the effective vertices
$\Gamma_{\{(f)\}P}^{c}$. In the NLLA the representation takes the
form~\cite{FF98}:
$$
\Phi_{P^{\prime }P}^{( {\cal R},\, \nu )} \left(
\vec{q}_{R};\,\vec{q};\, s_{0}\right) =\langle cc^{\prime
}|\hat{\cal {P}}_{\cal R}|\nu \rangle\int \frac{ds_{_{PR}}}{2\pi}
\theta \left( s_{_{\Lambda}}-s_{_{PR}}\right) \sum_{\left\{
f\right\} }\int \Gamma _{\left\{ f\right\} P}^{c}(q_{R})\left(
\Gamma _{\left\{ f\right\} P^{\prime }}^{c^{\prime
}}(q_{R}^{\prime })\right) ^{\ast }\left(
\frac{s_{0}}{\vec{q}_{R}^{~2}}\right) ^{\frac{1}{2}\omega \left( -\vec{q}%
_{R}^{~2}\right) }
$$ \vspace*{-1mm}
\begin{equation}
\times  \left( \frac{s_{0}}{\vec{q}_{R}^{~\prime
2}}\right)^{\frac{1}{2}\omega \left( -\vec{q}_{R}^{~\prime
2}\right) }d\rho _{f}-\frac{1}{2}\int \frac{d^{D-2}r}{\vec{r}^{~
2} \vec{r}^{~\prime 2}}\Phi _{P^{^{\prime }}P}^{\left( {\cal R}
,\, \nu \right) {\scriptsize{B}}}\left( \vec{r},\vec{q}\right)
{\cal K}_{r}^{\left( {\cal R}\right) \scriptsize{B}} \left(
\vec{r},\vec{q}_{R}; \vec{q}\right) \, {\rm ln}\left(
\frac{s_{\Lambda }^{2}} {( \vec{r}-\vec{q}_{R})^2 s_{0}}\right)
\, . \label{z19}
\end{equation}
Here $s_{_{PR}}=\left( p_{P}-q_{R}\right) ^{2}$ is the squared
particle-Reggeon invariant mass.   The argument $s_{0}$ in the
impact factor  shows that it depends on the energy scale $s_{0}$
of the Mellin transformation in (\ref{5}). Of course, this
dependence disappears in the total expression for the amplitudes
(\ref{5}). The Born (LLA) impact factors $\Phi _{P^{^{\prime
}}P}^{\left( {\cal R}, \, \nu \right)\, \scriptsize{B}}$ are
given by the first term in the r.h.s. of (\ref{z19}) taken in the
Born approximation. Note that for the Born case the integral over
$s_{_{PR}}$ in (\ref{z19}) is convergent, so that the parameter
$s_{_{\Lambda }}$ does not play any role. In  NLA the
independence of the impact factors on $s_{_{\Lambda}}$ is
guaranteed  by the factorization properties of the  effective
vertices $\Gamma_{\{(f)\}P}^{c}$~\cite{FF98}.

The non-forward impact factors of gluon and quark were calculated
at NLO in \cite{FFKP00_G} and \cite{FFKP00_Q} correspondingly for
arbitrary ${\cal R}$. Their general form is rather complicated,
so that we present here these impact factors only for the
antisymmetric octet representation (we omit the superscript
${\cal R} = 8-$, since we consider below only this
representation):
$$
\Phi^a_{G^\prime G}(\vec q_1, \vec q,
s_0)=\frac{-ig^2\sqrt{N}}{2}T^a_{G^\prime G}\left[
\delta_{\lambda_G, \lambda_{G^{\prime }}}\biggl\{ 1 + \frac
{\omega^{(1)}(t)}{2}\biggl[ {\tilde K_1} + \left( \left(
\frac{\vec q_1^{\:2}}{\vec q^{\:2}} \right)^\epsilon + \left(
\frac{\vec q_1^{\:\prime \:2}}{\vec q^{\:2}} \right)^\epsilon - 1
\right) \right.
$$
$$
\times \biggl( \frac{1}{2\epsilon} + \psi(1+2\epsilon) -
\psi(1+\epsilon)+
\frac{11+7\epsilon}{2(1+2\epsilon)(3+2\epsilon)} -
\frac{n_f}{N}\frac{(1+\epsilon)}{(1+2\epsilon)(3+2\epsilon)}
\biggr)
$$
$$
+ \ln\left( \frac{s_0}{\vec q^{\:2}} \right) +
\frac{3}{2\epsilon} + 2\psi(1) - \psi(1+\epsilon) -
\psi(1+2\epsilon)
$$
$$
-\frac{9(1+\epsilon)^2+2}{2(1+\epsilon)(1+2\epsilon)(3+2\epsilon)}
+\frac{n_f}{N}\frac{(1+\epsilon)^3+\epsilon^2}{(1+\epsilon)^2(1+2\epsilon)
(3+2\epsilon)} \biggr] \biggr\}
$$
\begin{equation}\label{37}
\left.- \frac{\delta_{\lambda_G,
-\lambda_{G^{\prime}}}\epsilon\omega^{(1)}(t)}
{2(1+\epsilon)^2(1+2\epsilon)(3+2\epsilon)}\left( 1 + \epsilon -
\frac{n_f}{N} \right)\right]\,,
\end{equation}
$$
\Phi^a_{Q^\prime Q}(\vec q_1, \vec q, s_0)
=\frac{-ig^2\sqrt{N}}{2}T^a_{Q^\prime Q}\delta_{\lambda_Q,
\lambda_{Q^{\prime }}}\biggl( 1 + \frac{\omega^{(1)}(t)}{2}
\biggl[ {\tilde K_1} + \left( \left( \frac{\vec q_1^{\:2}}{\vec
q^{\:2}} \right)^\epsilon + \left( \frac{\vec q_1^{\:\prime
\:2}}{\vec q^{\:2}} \right)^\epsilon \right) \biggl\{
\frac{1}{2\epsilon}
$$
$$
+ \psi(1+2\epsilon) - \psi (1+\epsilon)+
\frac{11+7\epsilon}{2(1+2\epsilon)(3+2\epsilon)} -
\frac{n_f}{N}\frac{(1+\epsilon)}{(1+2\epsilon)(3+2\epsilon)}
\biggr\}
$$
\begin{equation}\label{31}
+ \ln\left( \frac{s_0}{\vec q^{\:2}} \right) + 2\psi(1) -
2\psi(1+2\epsilon) - \frac{3}{2(1+2\epsilon)} - \frac{1}
{N^2}\left( \frac{1}{\epsilon} -
\frac{(3-2\epsilon)}{2(1+2\epsilon)} \right) \biggr] \biggr)\,,
\end{equation}
where $T^a_{P^\prime P}$ are the matrix elements of the colour
group generators in corresponding representations, $\lambda_P$ are
the particle helicities and the function ${\tilde K_1}$ has the
following integral representation:
\begin{equation}\label{32}
{\tilde K_1} =
\frac{(4\pi)^{2+\epsilon}\Gamma(1+2\epsilon)\epsilon\left( \vec
q^{\:2} \right)^{-\epsilon}}
{4\Gamma(1-\epsilon)\Gamma^2(1+\epsilon)}\int\frac{d^{D-2}k}{(2\pi)^{D-1}}\ln\left
( \frac{\vec q^{\:2}} {\vec k^{2}} \right)\frac{\vec
q^{\:2}}{(\vec k - \vec q_1)^2(\vec k - \vec q_1^{\:\prime})^2}\;.
\end{equation}
The result of the integration of (\ref{32}), in form of expansion
in $\epsilon$, can be found in \cite{FFKP00_G}.

Of course, the most interesting is the impact factor for virtual
photon, since it is the impact factor of physical particle which
can be calculated in perturbative QCD. In the leading order it
differs only by a numerical factor from the QED case, which is
known long ago. Unfortunately, at NLO this impact factor is not
yet calculated, although some steps in this direction are done
\cite{FM99},\cite{BGQ}.

\section{"Bootstrap" of the gluon Reggeization}

The "bootstrap" condition for the kernel derived in
Ref.~\cite{FF98} has the form
\begin{equation}
\int\frac{d^{D-2}q_1}{\vec q_1^{\:2}\vec q_1^{\:\prime \:2}}\left[
\int\frac{d^{D-2}q_2}{\vec q_2^{\:2}\vec q_2^{\:\prime \:2}}
 {\cal K}^{\left( 8-\right) }\left( \vec q_1,\vec
q_2;\vec q\right) - \omega(t) \right] = 0~,
\label{28}
\end{equation}
where ${\cal K}^{\left(8-\right)}\left(\vec q_1,\vec q_2;\vec
q\right)$ is the kernel of the non-forward BFKL equation in the
gluon (antisymmetric octet) channel, $\omega \left(t\right)
=\omega^{\left(1\right)}\left(t\right)
+\omega^{\left(2\right)}\left(t\right)$ is the deviation of the
gluon Regge trajectory from unity and $t=-\vec q^{\:2}$.

The fulfillment of this condition for the quark contribution to
the kernel was shown in \cite{FFP} at arbitrary $D$. In the
one-loop approximation the trajectory is purely gluonic (see
(\ref{8})). The quark contribution to the trajectory appears at
the two-loop level  and is given by the terms proportional to
$n_f$ in  (\ref{18}), (\ref{19}). The kernel ${\cal K}^{\left(
8\right) }\left( \vec q_1,\vec q_2;\vec q\right) ,$ according to
(\ref{7}), is expressed through the trajectory and the "real"
part. The quark piece of the latter is given by (\ref{23}). Using
this equation together with (\ref{18}), (\ref{19}) for the quark
contribution to the trajectory, we obtain
\[
\int \frac{d^{D-2}q_2}{\vec q_2^{\:2} \vec q_2^{\,\prime ~2}}
{\cal K} ^{\left( 8-\right) Q}\left( \vec q_1,\vec q_2;\vec
q\right) = {\Large g^4n_fN\frac 1{(2\pi )^{D-1}}\frac 1{(4\pi
)^{2+\epsilon }}\frac{ \Gamma (1-\epsilon )}\epsilon
\frac{[\Gamma (2+\epsilon )]^2}{\Gamma (4+2\epsilon )} }
\]
\[
\times \int \frac{d^{D-2}q_2}{\vec q_2^{\:2}}\left\{ {\Large \frac{\vec
q_1^{\:2}}{{\vec k}_{}^2}}\left( (\vec q_1^{\:2})^\epsilon +(\vec
q_1^{\:\prime \:2})^\epsilon -(\vec q_2^{\:2})^\epsilon -(\vec q_2^{\:\prime
\:2})^\epsilon \right) \right.
\]
\begin{equation}
\left. \ +\frac{{\Large \vec q^{\:2}}}{\vec q_2^{\:\prime \:2}}\left( (
{\Large \vec q^{\:2})}^\epsilon -(\vec q_1^{\:2})^\epsilon -(\vec
q_2^{\:2})^\epsilon \right) +\;\;(\vec q_1\longleftrightarrow \vec
q_1^{\:\prime })\right\} ~,  \label{29}
\end{equation}
where ${\vec k=}\vec q_1-\vec q_2\:$, $\vec q_i^{\:\prime }=\vec
q_i-\vec q$. Putting (\ref{29}) in (\ref{28}) and using
(\ref{18}), (\ref{19}), it is easy to verify that the "bootstrap"
equation (\ref{28}) is satisfied.

To demonstrate the fulfillment of the "bootstrap" condition for
the gluon piece of the kernel is not so simple. In \cite{FFK00} it
was done for the terms in the gluon trajectory non-vanishing for
$\epsilon \rightarrow 0$. Even for this purpose the knowledge of
the "real" part of the kernel ${\cal K}_r^{G\,(8-)}(\vec q_1,\vec
q_2; \vec{q})$ (\ref{24}) is not sufficient on account of the
singularity of the integration measure in (\ref{28}) at zero
momenta of scattered Reggeons. Fortunately, ${\cal
K}_r^{G\,(8-)}(\vec q_1,\vec q_2; \vec{q})$ turns into zero at
this points (see (\ref{16})),  so that at first sight these points
could not bring additional singularities in $\epsilon $. It would
be so if integration over only one momentum was performed. But in
the region where the momenta of two Reggeons
(let us say, $\vec{q}_{1}^{\,}$ and $%
\vec{q}_{2}^{\,}$) turn into zero simultaneously, being of the
same order, ${\cal K}_r^{G\,(8-)}(\vec q_1,\vec q_2; \vec{q})$
does not vanish (as it can be easily seen in the example of the
Born case (\ref{9})). Therefore, these regions can give additional
singularities in $\epsilon $; moreover, since integration over
these regions
leads to singularities, an expansion of, let us say, $(\vec{q}%
_{1}^{\,2})^{\epsilon }$ and $(\vec{q}_{2}^{\,2})^{\epsilon }$ is
not possible anymore, so that we need to know about the kernel
more than it is given by  (24). This problem was solved in
\cite{FFK00} and the fulfillment of the "bootstrap" equation
(\ref{28}) was proved for the terms in $\omega^{(2)}$
non-vanishing for $\epsilon \rightarrow 0$. After the calculation
of the RRG vertex at arbitrary $D$ \cite{FFP00} examination of
the "bootstrap" condition became possible at any you like
dimension. This question is under consideration now.

The "bootstrap" conditions for the impact factors derived in
\cite{FF98} are:
\begin{displaymath}
ig\frac{\sqrt{N}t}{\left( 2\pi \right) ^{D-1}}\int \frac{d^{D-2}q_1}{%
\vec{q}_1^{~2}\vec{q}_1^{~\prime 2} }\Phi
_{P^{^{\prime }}P}^{c}\left( \vec{q}_1%
^{~\prime };\vec{q};s_{0}\right)
\end{displaymath}
\begin{equation}
=\Gamma _{P^{\prime }P}^{c }(q) \omega ^{\left( 1\right) }\left(
t\right) +\Gamma _{P^{\prime}P}^{c\left(
{\scriptsize{B}}\right)}(q) \frac{1}{2}\left[\omega ^{\left(
2\right) } \left( t\right)-\left(\omega ^{\left( 1\right) } \left(
t\right)\right)^2 {\rm ln}\left(
\frac{\vec{q}^{~2}}{s_{0}}\right) \right] \ . \label{z34}
\end{equation}
The fulfillment of these conditions for the gluon and quark
impact factors was demonstrated  in \cite{FFKP00_G} and
\cite{FFKP00_Q} correspondingly  at arbitrary $D$.

More strong  conditions than (\ref{28}),(\ref{z34}) were suggested
in \cite{BV} (from  another point of view  they were considered in
\cite{FFKP00}). This suggestion is based on the requirement of
Reggeization of the unphysical particle-Reggeon scattering
amplitude with colour octet in the $t$-channel and negative
signature.  It was shown in \cite{BV}, \cite{FFKP00} that the
strong bootstrap conditions are satisfied in the case of the
gluon and quark impact factors. It is rather intriguing, because
their role is not yet clear.

\end{document}